\documentclass[prl,twocolumn,showpacs]{revtex4}
\usepackage{graphicx}
\usepackage{amsmath}

\begin{document}
\title{Experimental Quantum Coin Tossing}
\author{G. Molina-Terriza$^1$\footnote{Present address: Institut de Ci\`encies Fot\`oniques, Barcelona, SPAIN}, A. Vaziri$^1$\footnote{Present address: Atomic Physics Division, National Institute of Standards and 
Technology, Gaithersburg, Maryland 20899, USA}, R. Ursin$^1$ and
A. Zeilinger$^1,^2$}

\affiliation{$^1$Institut f\"ur Experimentalphysik, Universit\"at
Wien, Boltzmanngasse, 5, A-1090, Vienna, Austria\\ $^2$Institute for Quantum Optics and Quantum
Information, Austrian Academy of Sciences}

\begin{abstract}

In this letter we present the first implementation of a quantum coin tossing
protocol. This protocol belongs to a class of ``two-party'' cryptographic problems, where the communication partners distrust each other. As with a number of such two-party protocols, the best implementation of the quantum coin tossing requires qutrits. In this way, we have also performed the first complete
quantum communication protocol with qutrits. In our experiment the two
partners succeeded to remotely toss a row of coins using photons entangled in the
orbital angular momentum. We also show the experimental bounds of a possible cheater and
the ways of detecting him. 
\end{abstract}

\date{\today}


\maketitle

In the original ``coin tossing'' protocol, Alice and Bob had just
divorced and didn't want to ever see each other again but they
had to decide who kept the dog \cite{Blum81}. As they didn't
trust any third party as a referee, they agreed to toss a coin to
decide. How could, let's say, Bob be sure, that Alice is honest
when she said "It was tails... you lost" if he could not see the
outcome of the toss?. This simple protocol is at the heart of
other more complicated cryptographic problems, like mail
certification, remote contract signing and mental poker. Also,
this protocol belongs to a set of cryptographic problems where we
don't distrust a third party who can eavesdrop our secret, but
the problem is to control the information that the two
communication partners share. This set of protocols are usually
called ``post cold war'' protocols \cite{GottesmanLo00}. Other examples of this kind
of problems are the ``bit commitment'' protocol or the
computation of a function where the inputs are distributed.  

In the last years this kind of protocols have received a lot of 
attention from both the cryptographic and quantum 
information communities. Although the perfect security of some
of the ``two-party'' protocols seems impossible \cite{LoChau97,Mayers97, LoChau98FP}, 
it is yet unclear which bounds can be imposed on the security. There have been also some
speculation about the possibility that quantum mechanics can be 
derived only from purely quantum information postulates, stating the
possibility and impossibility of, respectively, quantum key distribution and
quantum bit commitment \cite{BitCommit2}.

In the case of ``coin tossing'', a set of  solutions to this problem works 
in the following way:
Alice throws the coin, locks it in a ``box'' and sends it to Bob.
Bob has the proof that the coin was thrown, but cannot see the
actual result. Bob makes his bet and, upon receiving his bet,
Alice sends the key to Bob, so that he is able to unlock the
result \cite{BitCommit}. Up to now, there is no proof
that the ``boxes'' used in the classical implementations of this
protocol (for example, one-way functions), are truly impossible
to unlock by Bob, or cannot be modified by Alice (see Fig. 1(a)).
In general, there is no classical protocol which allows
unrestricted security against cheating for the ``coin tossing''
protocols. On the other hand, using quantum mechanics, it is possible
to at least limit the ability of any party to cheat.

In quantum coin tossing \cite{Ambainis01a,BB84,LoChau98,Aharanov00,SpekRudo}, we
replace the ``box'' by a quantum state. Alice chooses one among a
series of non-orthogonal states and sends it to Bob. Each of the
states encodes the result of the throw of the coin. In this way,
without previous knowledge, Bob cannot know with certainty which
of the states he possesses. At this point, Bob makes his bet. To
``unlock'' the state, Alice only has to tell Bob which was the
state she sent and then he can measure it in an orthogonal basis
to check that Alice is honest. If Bob's measurement corresponds
with Alice's predicted state, the protocol is a success.
Otherwise, Alice and Bob would consider the throw as a
``failure'' and would disregard it. This coin tossing scheme
limits the probability of a cheater to succeed. We will show
that, in reality, cheating is actually detected when the
``failures'' in a row of throws increase over the statistical
errors.

Our implemented protocol is based on a proposal by Ambainis \cite{Ambainis01a}
which uses three-dimensional quantum
states (also called ``qutrits''). Up to now, all the devised
protocols using qubits allow a theoretical higher probability for
the cheater to win \cite{SpekkensRudolph02}. In this sense the
protocols using qutrits are better suited for this particular
problem. This raises an interesting question about which other
problems can also be more efficiently solved with higher
dimensional states \cite{Whitegroup}.

The series of states that Alice can send and the correspondent
throw of the coin are presented in Table 1 \cite{weakct}. Alice's
states are divided into two sets, each of them containing two
orthogonal states. States of one set have a non-vanishing
projection onto states of the other set. For this reason, Bob
needs two different measuring bases in order to determine the
state of each possible photon sent by Alice. In Table 1, we also
show the elements of Bob's bases associated with each set. Note
that every basis contains, besides the states of the
corresponding set, a third state orthogonal to them, which in our
case is either $|2\rangle$ (set $1$) or $|1\rangle$ (set $2$).
These additional states are crucial for increasing the chances to
detect cheating, as we will show below. Ambainis
\cite{Ambainis01a} demonstrated that the maximum probability that
one of the partners biases the result without being noticed is
$25\%$. Remember that, in any coin tossing, each party needs to
cheat only in $50\%$ of the throws, because that's the
probability of losing, which limits the theoretical probability
of a cheater to win to $75\%$.

\begin{table}
\begin{tabular}{|c|c|c|c||c|c|}
    \hline Set & Label & Alice's States & Coin & Bob's Bases & Label \\ \hline \hline
    & & & & & \\
  $1$ & A$11$ & $(|0\rangle+|1\rangle)/\sqrt{2}$ & Heads ($1$) & \raisebox{2ex}
[2ex]{$(|0\rangle+|1\rangle)/\sqrt{2}$} &
  \raisebox{2ex}[2ex]{B$11$}
  \\ \cline{5-6}
    & & & & & \\ \cline{1-4}
    & & & & \raisebox{2ex}[2ex]{$(|0\rangle-|1\rangle)/\sqrt{2}$} & \raisebox
{2ex}[2ex]{B$12$} \\ \cline{5-6}
 $1$ & A$12$ & $(|0\rangle-|1\rangle)/\sqrt{2}$ & Heads ($1$) & &\\
    & & & &  \raisebox{2ex}[2ex]{$|2\rangle$} & \raisebox{2ex}[2ex]{B$13$}  \\
\hline \hline

    & & & & & \\
  $2$ & A$21$ & $(|0\rangle+|2\rangle)/\sqrt{2}$ & Tails ($0$) & \raisebox{2ex}
[2ex]{$(|0\rangle+|2\rangle)/\sqrt{2}$} &
  \raisebox{2ex}[2ex]{B$21$}
  \\ \cline{5-6}
    & & & & & \\ \cline{1-4}
    & & & & \raisebox{2ex}[2ex]{$(|0\rangle-|2\rangle)/\sqrt{2}$} & \raisebox
{2ex}[2ex]{B$22$} \\ \cline{5-6}
 $2$ & A$22$ & $(|0\rangle-|2\rangle)/\sqrt{2}$ & Tails ($0$) & &\\
    & & & &  \raisebox{2ex}[2ex]{$|1\rangle$} & \raisebox{2ex}[2ex]{B$23$}  \\
\hline
\end{tabular}
\caption{Here we show the four different states sent by Alice and
the bases used by Bob to properly characterize the incoming
photon. The Alice's states are divided into two sets of two
states. Each set represents a particular side of the coin. Bob
uses two basis, corresponding to Alice's states, each expanded by
one further orthogonal state. The label of the states eases their
recognition in Fig. 1}
\end{table}

In our experiment Alice and Bob agreed to build the following
set-up to implement the protocol (See Fig. 1(b)). Alice possess a
source of orbital angular momentum entangled photons 
\cite{Mair01a,Molina-Terriza02}. She keeps
one photon of the pair and sends the other one to Bob. When she
projects her photon onto one of the four states in Tab. 1 and
detects it, she knows that she is sending to Bob a triggered
photon \cite{triggered} carrying an orbital angular momentum
qutrit. Together with the photon, she sends a signal to Bob,
telling him that a coin has been thrown and so, the corresponding
entangled photon with the right state is being sent to him. Once
Bob receives the signal, he sends his bet to Alice. Now she can
tell Bob which was the state. Using this information, Bob can
measure the state of the photon and verify the honesty of Alice.

The set-up consists of a $351$nm wavelength Argon-ion laser pumping 
a $1.5-$mm-thick BBO ($\beta$-barium-borate) crystal cut for Type I 
phase matching condition. The crystal is positioned such as to produce
down-converted pairs of identically polarized photons at a
wavelength of $702$nm emitted at an angle of $4^\circ$ off the
pump direction. These photons are directly entangled in the
orbital angular momentum degree of freedom. One of the photons is
sent to Bob, meanwhile the other remains on Alice's side. With a
series of beam-splitters, whose reflectances were chosen so that
the photons were equally distributed among the different paths,
both parties direct their photons probabilistically to a series of
holograms and single mode fibers, prepared to project the photon
onto a certain state \cite{Ali_JOB}. Alice's and Bob's states are
presented in Table 1.

Bob's measurements were performed in the following way: Bob's
photons were directed randomly to a projection onto one of the
six possible states. The photons going to the wrong basis were
discarded.

Once the set-up was built, Alice and Bob decided to try their
coin-tossing protocol with a row of throws. In this experiment
they obtained $50\%$ heads ($1$) and $44\%$ tails ($0$). As Bob's
guesses were random, he won in half of the throws. The overall
failures in the protocol represented around $6\%$ of the throws,
intrinsic to the setup. In Fig. 2 (a)-(b) we present the
experimental probabilities of Bob's measurements, for two
different states sent by Alice.

In Fig. 3(a), we show a set of the actual throws. Every square of
the image represents a throw of the coin. The color of every
square represents the result of the throw, as Alice communicates
it to Bob (black is tails, white is heads). Red (gray) means that the
throw was a failure.

In order to explore the limits of the implementation Alice
decided to cheat Bob. We want to remark that it is a harder
problem to devise a cheating procedure, than to prepare the honest
protocol. In our case, the best way we could find for Alice to
cheat was the following one. Alice always sends a random
symmetric  mixture of the state $(|0\rangle-|1\rangle)/\sqrt{2}$
and the state $(|0\rangle+|2\rangle)/\sqrt{2}$, which is a random
mixture of being heads or tails \cite{cheat}. When Bob makes his
bet about the state, Alice always tells him that he lost, and
then, Bob has to measure in the corresponding basis. For example,
if Bob says that it was tails, Alice's answer is that the state
was heads ($(|0\rangle-|1\rangle)/\sqrt{2}$), and Bob measures in
this basis. It is an easy task to check that Alice will win in
$63.5\%$ of the throws, which is below the theoretical maximum of
$75\%$ \cite{Ambainis01a,SpekkensRudolph02}. This strategy
resembles a biased coin which can be flipped even after Bob made
his choice.

In order to force Alice to cheat, we changed states A$12$ and
A$21$ to be identical to states number A$22$ and A$11$,
respectively. Alice does not need to keep a record of which state
is being sent. In this way she sends a probabilistic mixture of
the two states, which represent correspondingly heads and tails.
 Her strategy is to tell Bob that the state sent was exactly the 
 opposite of Bob's bet.

In the experimental implementation, of course, it is even harder
for Alice to win, because she cannot turn off the statistical
errors which also happen in the honest protocol. In our case, we
found that the number of failures, when Alice cheats in this way,
was a $46\%$ of the throws. In Fig. 2 (c)-(d), one can check Bob's
results for the two different states that Alice claimed she was
sending. Also, we present the actual row of throws in Fig. 3(b),
with the same coding colors than Fig. 3(a). In this case, by
comparing the results presented in images (a) and (b) it was very
easy for Bob to discover that in (b) the protocol was not followed
in the honest way. Also, it was a little bit suspicious to him
that in all the proper throws of Fig. 3(b), Alice won.

At this point, let us turn back to Fig. 2(c)-(d). We want to note
that the state in which most of the failures go is precisely the
one which is outside the plane defined by the elements of Alice's
set. It is unclear whether other cheating procedures would give
rise to the same result but, at least in this case, the use of a
three dimensional space is necessary in order that Bob can detect
Alice cheating.

Once the subject was clarified between Alice and Bob, he decided
to build a table where he could infer the amount of dishonest
throws, given the number of failures during the protocol (Fig.
3(c)).

Before finishing, we would like to discuss a few details from
this particular implementation. In a proper protocol, the
detection of the photon by Bob should be delayed until he can
send his bet to Alice. In our set-up, it was very difficult to
prepare such a delay and so, we simulated it by software. A more
realistic implementation should include an optical delay with,
for example, a couple of parallel mirrors.

Another difference with respect to an ideal implementation is that
both Alice and Bob could not deterministically project their
photon onto a given state. Although it is clear that this is not a
problem for an honest Alice, who chooses at random which state to
send among the possible four, it might present a security hazard
when one of the parties cannot be trusted. A closer look to the
set-up shows that this is not the case.

If Bob's photon goes to the wrong projection, Bob just loses the
photon and has to ask Alice to send it again. Alice cannot take
profit of this effect, because she never knows whether the photon
she sends is going to the right projection or not.

On the other hand, Bob could try to use this probabilistic
behavior in his favor, just by not acknowledging a right
measurement of a lost bet. But then Alice would notice an abnormal
increase of lost photons and would stop the communication, in the
same way as an increase of ``failures'' would be the trademark of
a cheater.

In conclusion, we have experimentally demonstrated a ``quantum
coin tossing'' protocol. To our knowledge, this is the first
``two-party'' communication protocol which is solved using the
laws of quantum physics to encode the communication. It is worth
mentioning that, contrary to the usual ``key distribution''
protocols, in this case the information shared by Alice and Bob
is truly exchanged through the quantum states. Also, this
protocol is the first to be implemented, where the use of more
than two dimensions presents a clear advantage. Using our set-up
we could share a set of a few tens of thousand coin throws in a
few seconds among two parties. We also allowed one party to try to
cheat, which could be easily detected by a significant increase
of ``failures''. We could not find an optimal cheating procedure,
but we hope that this work triggers others, where the
possibilities of realistic dishonest parties are studied.

This work was supported by the Austrian
Sciences Foundation  (F.W.F) and the European Commission through
the Marie-Curie program and the RAMBO-Q project of the IST
program. Discussions with Markus Aspelmayer and Kevin Resch are
gratefully appreciated.

\newpage

\begin{figure}
\caption{a) Sketch of the implemented ``coin
tossing'' protocol: First step, Alice throws a coin, encodes it
in a quantum state and sends it to Bob. Second step, Bob's sends
his bet to Alice. Third step, Alice tells Bob which state it was
and Bob unlocks the result by measuring the state. b) Diagram of
the set-up used. Alice possesses a source of entangled photons.
Using beamsplitters, she projects probabilistically one of the
photons onto one of the four possible states shown in Table 1.
This state is transferred nonlocally to the other photon, which
is on its way to Bob. Bob's photon is projected randomly onto one
of the six possible elements of the two bases. Photons going to a
wrong basis are not considered.
\label{fig1}}
\end{figure}

\begin{figure}
\caption{Statistics of Bob's measurements. Red (gray) bars
correspond to ``failures'' of the protocol. Black (heads) and
white (tails) bars correspond to proper throws and indicate the
result of the tossing. (a)-(b) Alice is honest. We present only
two of the possible four states sent by Alice: (a) she sends the
state $|0\rangle+|1\rangle)/\sqrt{2}$ (Heads) and (b) she sends
the state $|0\rangle-|2\rangle)/\sqrt{2}$ (Tails). The errors in
this case are due to misalignments of the set-up and are
intrinsic to it. (c)-(d) Alice is cheating. She always sends a
mixture of two states. After Bob makes his bet, she decides which
state she must tell him: (c) she claims to send state
$|0\rangle-|1\rangle)/\sqrt{2}$, (d) she claims to send state
$|0\rangle+|2\rangle)/\sqrt{2}$. In this case, the errors are
clear indicators of the presence of a cheater. The difference of
the errors of cases (c) and (d) are mainly due to the fact that
Alice is not sending a perfect mixture of the two states.
\label{fig2}}
\end{figure}

\begin{figure}
\caption{(a)-(b) Two different ordered rows of
throws. Upper left corner: first throw, lower right corner: last
throw. Color code as in Fig. 2, Black: head, White: tail and
Red (gray): ``Failure''. (a) Bob and Alice are honest: Heads $50\%$,
tails $44\%$, failures $6\%$. The difference between heads and
tails due to different efficiencies of the detectors and the
failures due to imperfections of the set-up. (b) Bob honest, Alice
cheats. Heads $26\%$, tails $28\%$, failures $46\%$. One can
clearly see how the failures increase due to the fact that Alice
is cheating. (c) Probability of failure as a function of the
amount of cheating by Alice. Solid line: theory. Circles:
Experimental data.
\label{fig3}}
\end{figure}


\begin{thebibliography}{10}
\bibliographystyle{unsrt}


\bibitem{Blum81}
M. Blum, Coin Flipping by Phone. {\it CRYPTO} 1981, 11--15 (1981)

\bibitem{GottesmanLo00}
D. Gottesman. J.-K. Lo, {\it Physics Today} {\bf 53}, 22--27
(2000).

\bibitem{LoChau97} 
H.-K.Lo and H.F. Chau, {\it Phys. Rev. Lett.} {\bf 78}, 3410--3413 (1997)

\bibitem{Mayers97} 
D. Mayers, {\it Phys. Rev. Lett.} {\bf 78}, 3414--3417 (1997)

\bibitem{LoChau98FP} 
H.-K.Lo and H.F. Chau, {\it Fortschr. Phys.} {\bf 46}, 507--519 (1998)

\bibitem{BitCommit2} J. Bub {Foundations of Physics} {\bf 31} 735--756
(2001)

\bibitem{BitCommit} Explained in this way it is easy to see the
similarities between this protocol and a related one: bit
commitment. Nevertheless, bit commitment is a stronger protocol
than coin tossing. For a review on quantum bit commitment see
Ref. \cite{BitCommit2}

\bibitem{Ambainis01a}
A. Ambainis, A new protocol and lower bounds for quantum coin
flipping, {\it Proceedings of STOC}, 134--142 (2001).

\bibitem{BB84}
C. H. Bennett and G. Brassard, {\it Proceedings of IEEE International
Conference on Computers, Systems, and Signal Processing}, 175--179 (1984)

\bibitem{LoChau98} H.-K. Lo, H. F. Chau, {\it Physica D} {\bf 120}, 177 (1998).

\bibitem{Aharanov00} D. Aharonov, A. Ta-Shma, U. Vazirani, A. Yao,
{\it Proceedings of the 32nd Annual Symposium on Theory of
Computing 2000 (Association for Computing Machinery, New York,
2000)}, p. 705.

\bibitem{SpekRudo} R. W. Spekkens, T. Rudolph, {\it Phys. Rev. A}
{\bf 65}, 012310 (2001).

\bibitem{SpekkensRudolph02} R. W. Spekkens, T. Rudolph,
{\it Phys. Rev. Lett.} {\bf 89}, 227901 (2002).

\bibitem{Whitegroup}  N. K. Langford et al. quant-ph/0312072

\bibitem{weakct} For simplicity reasons, we have chosen to describe
a slightly different version of the original protocol \cite{Ambainis01a}. The
protocol described belongs to a class of protocols called Weak
Coin Tossing. In our set-up it is straightforward to implement
also the strong version. In this case, Bob makes his bet before
starting (either `0' or `1'). Alice encodes one bit $b$ in her
state following Table 1 and sends it. Bob sends a classical bit
$b^\prime$ to Alice. After checking that the protocol was
successful, the thrown of the coin is given by $b\oplus b^\prime$.

\bibitem{Mair01a}
A. Mair, A. Vaziri, G. Weihs, A. Zeilinger, {\it Nature} {\bf
412}, 313 (2001).

\bibitem{Molina-Terriza02} G. Molina-Terriza, J. P. Torres,
L. Torner, {\it Phys. Rev. Lett.} {\bf 88}, 013601 (2002).

\bibitem{triggered} G. Molina-Terriza {\it et al.} To be published 
in Phys. Rev. Lett. Available at http://arxiv.org/abs/quant-ph/0401183

\bibitem{Ali_JOB} A. Vaziri, G. Weihs, A. Zeilinger, {\it J. Opt. B:
Quantum Semiclass. Opt.} {\bf 4} S47 (2002)

\bibitem{cheat} Another strategy for Alice to cheat is as follows: she
just sends a single state, and no matter Bob's bet, she always
says that he lost. Although, the set-up is easier to implement,
in this case the probability of Alice to win depends on Bob's
choice.
\end{thebibliography}
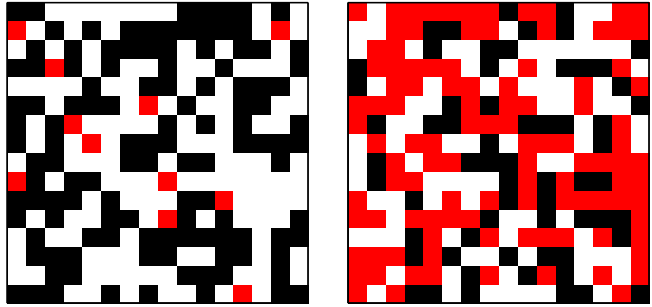
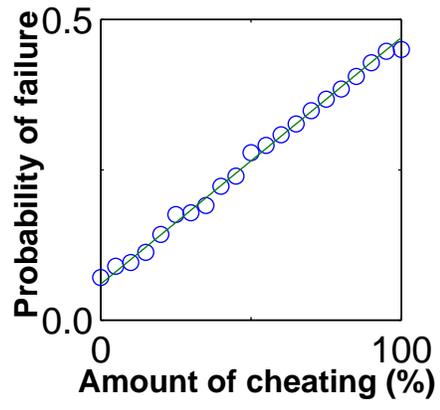
\end{document}